\documentclass[aps,prd,twocolumn,showpacs,amsmath,amssymb]{revtex4}
\usepackage{bm}
\usepackage{latexsym}
\usepackage{multirow}

\usepackage{color,graphicx}
\usepackage{dcolumn}
\usepackage{bm}
\usepackage{mathrsfs}
\usepackage[T1]{fontenc}

\newcommand{\be}{\begin{equation}}
\newcommand{\ee}{\end{equation}}
\newcommand{\ba}{\begin{eqnarray}}
\newcommand{\ea}{\end{eqnarray}}
\newcommand{\ban}{\begin{eqnarray*}}
\newcommand{\ean}{\end{eqnarray*}}

\begin{document}

\title{Bekenstein-Hawking entropy from Criticality}

\author{Swastik Bhattacharya}
\email{swastik@iisertvm.ac.in}
\affiliation{School of Physics, Indian Institute of Science Education and Research Thiruvananthapuram (IISER-TVM), Trivandrum 695016, India}

\author{S. Shankaranarayanan} 
\email{shanki@iisertvm.ac.in} 
\affiliation{School of Physics, Indian Institute of Science Education and Research Thiruvananthapuram (IISER-TVM), Trivandrum 695016, India}

\begin{abstract} 
Vacuum Einstein equations when projected on to a black hole horizon is analogous to the dynamics of fluids. 
In this work we address the question, whether certain properties of semi-classical black holes could
be holographically mapped into 
properties of (2 + 1)-dimensional fluid living on the horizon. In particular, we focus on the 
statistical mechanical description of the horizon-fluid that leads to Bekenstein-Hawking entropy. Within the 
paradigm of Landau mean field theory and existence of a condensate at a critical temperature, we 
explicitly show that Bekenstein-Hawking entropy  and other 
features of black hole thermodynamics can be recovered from the statistical modelling of the fluid.  We also 
show that a negative cosmological constant acts like an external magnetic field that induces order in the system 
leading to the appearance of a tri-critical point in the phase diagram.
\end{abstract}
\maketitle

\section{Introduction}

In 1970's, it was shown that the dynamics of Black-holes is formally analogous to thermodynamics 
\cite{BH-Thermo}. 
However, it was only with the discovery of Hawking radiation\cite{Hawking-1975}, that the precise 
expression of black-hole entropy was found. The laws of black-hole mechanics also came to be 
viewed widely as the laws of black-hole thermodynamics. Similarly, in  recent years, 
it has been shown that the dynamics of gravity, near the black-hole horizon, is analogous 
to the dynamics of fluids \cite{Shiraj,Paddy,Strominger}. (For earlier works, see Refs. 
\cite{Sakharov, Damour, Membrane}.)

As in the case of black-hole mechanics, the Fluid-Gravity correspondence is still formal. 
Currently, it is used as an operational tool to study a relatively simpler problem on one side 
(say, Fluid), use the correspondence and know about a much harder problem in the Gravity sector\cite{Shiraj}. 
In this work, we ask a different question related to the black-hole entropy: 
Can the Fluid-Gravity analogy provide a statistical mechanical description of the 
$(2+1)-$D fluid living on the black hole horizon that leads to Bekenstein-Hawking entropy and in general, the 
black-hole entropy?  If this can be achieved, then the Fluid-Gravity correspondence may be more than an 
analogy.

There have been attempts in the literature to obtain black-hole entropy from degrees of freedom 
of quantum fluids \cite{Laughlin,Dvali,stretchrzn}. Such attempts have proved unsatisfactory 
because the models cannot reproduce Black Hole Thermodynamics, in particular, the First Law and 
Bekenstein-Hawking entropy. In this work, we explicitly show that within the 
paradigm of Landau theory of phase transition and formation of a condensate at a 
critical temperature, Bekenstein-Hawking entropy including the correct pre-factor 
$\frac{1}{4}$ and other features of Black Hole Thermodynamics can be recovered from the 
statistical modelling of the fluids. 

The key assumption we make (based on the results of Ref. \cite{Skakala}), is the 
existence of a condensate that corresponds to the breaking of a continuous symmetry. We use Mean field theory 
to describe the phase transition leading to condensation. By construction, mean field theory includes only 
long wavelength fluctuations in the system and neglect the short-wavelength (high energy) 
fluctuations. This observation forms one of the crucial inputs in our analysis of reducing 
the symmetry group from continuous to discrete \cite{Mermin,Hohenberg,KadanoffKane}.  
(See Appendix \ref{RG} where we have shown this using the Path-integral technique.) The reduction of the 
symmetry group from continuous to discrete has the advantage that one can talk about a phase transition for 
the $2$-D systems\footnote{It is well known that long range order and phase transition cannot 
occur for continuous symmetries in $2$D \cite{Mermin,Hohenberg,Kadanoff}. A Bose condensate 
may form in $2$D, however, even if Phase Transition does not take place\cite{Nature,Cho}.}. 

Our approach predicts the existence of two phases of the Horizon-fluid system: Symmetric phase, 
in which, the black hole is in equilibrium with its surroundings. Non-symmetric phase, in which, 
black-hole is not in an equilibrium. It is in this phase, that the entropy is identical to Bekenstein-Hawking entropy. 
The negative cosmological constant gives rise to 
an external field, which favours long range order in the system. This is very similar to the introduction 
of an external magnetic field, which helps align the spins in a magnet. We show the existence of 
a tri-critical point in the phase diagram of the Horizon-fluid in the presence of a cosmological constant. 

In the next section, we develop the mean field theory formalism for the horizon-fluid and 
recover Bekenstein-Hawking entropy of Schwarschild black-hole in the non-symmetric phase. In section (III), we apply the 
formalism to Schwarzschild AdS and show the existence of a tri-critical point in the phase diagram of the 
Horizon-fluid. In section (IV), we discuss the physical significance of the order-parameter and 
give a physical understanding of the two phases. We end with conclusions and discussions. 

\section{Schwarzschild Black Hole}

To proceed beyond the macroscopic physics, we need to build a statistical model of the Horizon-Fluid system.
Using the fact that the black-holes are highly constrained,  the horizon-fluid can be modeled as a 
strongly correlated system. Further evidence for this 
has shown up recently where the authors have modeled the Horizon-fluid for a Schwarzschild black hole by 
a collection of massless Bose particles \cite{Skakala}. An interesting feature of the model is that the 
constraints force the Bose particles  to populate {\sl only} the ground state. 
(See Appendix (B) for the correspondence between the black-hole horizon as a null fluid and a Bose gas.)  
This strongly suggests the occurrence of a phase transition. With this insight, we make the following 
assumptions:
\begin{enumerate}
 \item  There is a temperature $T_c$ (critical temperature), at which, all the $N$ microscopic d.o.f. on  the 
 horizon form a condensate. 
%
\item The system always remains close to the critical point, where the phase transition takes place. 
\end{enumerate}

An immediate consequence of these assumptions is the deduction of the relation between $N$ and $A$ 
(see Appendix B). Since the system forms a condensate at $T_c$, nearly all the microscopic d.o.f.s would 
be 
in the ground state near the critical point. As there is only one scale in the problem, the total energy of 
the system can be expressed in the form, $E= N \alpha T$, where, $\alpha$ is a constant. Using 
the constraints \eqref{AE} and \eqref{ET},  we get
\begin{equation}
 N= \frac{A}{2\alpha}, \label{NA}
\end{equation}
We model the horizon-fluid system using mean field theory\cite{Landau, Stanley,Kadanoff}. 
The order parameter for a collection of particles, which forms a condensate at a certain transition 
temperature, is the wave function for the state ($\psi$) whose modulus is equal to the number density of 
particles $\rho$ i.e. $\psi \propto \sqrt{\rho}$. 

For mathematical simplicity, we assume the black hole horizon-fluid system to be homogeneous. Hence, 
the order parameter is given by:
\begin{equation}
 \psi= \sqrt{\kappa}\sqrt{N}. \label{psi}
\end{equation}
where $\sqrt{\kappa}$ contains the phase information. Following points need to be noted regarding the 
order parameter: (i) $\psi$ has a continuous $U(1)$ symmetry which will be broken beyond the critical point. 
(ii) The phase part of the order parameter presumably comes from the quantum dynamics of the microscopic 
d.o.f. on the horizon that are governed by high energy modes. This implies that the fluctuations in the 
phase part of $\psi$ occur at much smaller length scales than the fluctuations in the amplitude of $\psi$. 
Using the Renormalization Group analysis (details in Appendix \ref{RG}), we can rewrite the free energy 
of a macroscopic black hole in terms of the relatively low energy d.o.f. The only change that occurs is 
that the order parameter in the free energy can be treated as real. This reduces the symmetry group to 
$\mathbf{Z_2}$, a discrete one \cite{Domany1,Rottman,Domany2}. 
Henceforth, $\sqrt{\kappa}$ would be taken to be real valued. To distinguish the $\mathbf{Z_2}$ symmetric 
theory from the continuous one,  we define a new real valued order parameter:
\begin{equation}
 \eta= \sqrt{\kappa}\sqrt{N}, \label{eta}
\end{equation}
where, $\sqrt{\kappa}$ is now taken to be real valued. Of course, now, the physical significance of the order 
parameter could no longer be supplied directly from the microscopic model. Nonetheless, as we shall 
see later, it is possible to give a physical interpretation of $\eta$. 

Following Landau-Lifshitz \cite{Landau} (Section \S143), it will be better suited to write down the Mean field theory 
in terms of the Thermodynamic potential, $\Phi$, where the independent variables are $T$ and the chemical potential $\mu$\footnote{In 
\cite{Landau}, the Gibbs Free Energy is denoted by $\Phi$, whereas the Thermodynamic Potential mentioned here is denoted by 
$\varOmega$.} i .e., $ \Phi=-P \, A $. Expanding $\Phi$ about $T_c$, we have 
\begin{equation}
 \Phi= \Phi_0+a(P)(T-T_c)\eta^2+B(P)\eta^4. \label{phiLT}
\end{equation}
where $a(P)$ and $B(P)$ are unknown phenomenological functions.  Using \eqref{eqs}, \eqref{eta} and constraint \eqref{NA}, 
we get, 
\begin{equation}
   -TA= 4{\Phi_0+ \kappa a(P)(T-T_c)\frac{A}{2\alpha}+\kappa^2B(P)(\frac{A}{2\alpha})^2}
 \end{equation}

 Matching the coefficients of $A$ on both sides, we have, 
 \begin{equation}
  a= -\frac{\alpha}{2\kappa}, \label{a}
 \end{equation}
 which shows that $a$ is a negative number.  We also have the second mapping constraint, using \eqref{a},
  \begin{equation}
 \Phi_0(P,T)+\frac{1}{4}T_cA+ B(P)(\frac{A}{2\alpha})^2=0. \label{2mapc1}
\end{equation}

It is important to note that $\eta=0$ is the symmetric phase and $\eta \neq 0$ is the asymmetric phase 
\cite{Landau}. In our case, since $a < 0$, this implies that 
the system is in the symmetric phase for $T<T_c$  and asymmetric phase for $T>T_c$. Similar behaviour is 
exhibited in the case of Kosterlits-Thouless transition in 2-D systems \cite{Domany2}. 

In the asymmetric phase, the order parameter $\eta$ has the value for which, the Thermodynamic Potential is 
minimum. The minimisation of the Thermodynamic potential with respect to $\eta$ gives the condition
\begin{equation}
 \eta^2= \frac{a(T_c-T)}{2B} \Rightarrow \kappa N= \frac{a(T-T_c)}{2B}
  \label{minima}
\end{equation}

Using \eqref{NA} and \eqref{a}, this can be expressed as 
\begin{equation}
 \frac{(T-T_c)}{2B}= \frac{\kappa^2A}{\alpha^2}. \label{minima1}
 \end{equation}

Denoting the entropy of the system in 
the symmetric and  asymmetric phase by $S_0$ and $S_0+\Delta S$, respectively, we have 
\begin{equation}
 \Delta S= -\frac{\partial\Phi}{\partial T}= \frac{a^2}{2B}(T-T_c). \label{S1}
\end{equation}
 From \eqref{a} and \eqref{S1}, we get,
 \begin{equation}
  \Delta S= \frac{A}{4}. \label{S}
 \end{equation}
 This is one of the main results of the paper and we would like to discuss its importance: 
 First, in the semi-classical regime, the area $A$ of a macroscopic black hole is large implying that 
 $\Delta S$ is  very large. If $S_0$ is taken to be small compared to this, then we can approximate the 
 entropy $S$ 
 of the black hole-fluid system in the asymmetric phase to be $\frac{A}{4}$. This is the same as 
 given by Bekenstein-Hawking entropy of the black hole. Second, this analysis shows 
that the entropy calculated in Ref. \cite{Skakala} is the entropy of the system in the symmetric 
phase, which does not correspond to the black-hole entropy. Third, the specific heat cannot be 
defined in the standard way here as the black hole-fluid system is a one parameter system. However, 
it can be shown that  $\frac{dQ}{dT} < 0$. 
  
 \section{AdS-Schwarzschild Black hole}
 
$\Lambda$ can be incorporated in the mean field theory by introducing an external field $(h)$ that couples 
to the system. This is similar to switching on a magnetic field in a paramagnet-ferromagnet system. Later, we 
show that the external field is related to $\Lambda$. 

Using the relation $\Phi = - P\, A$, we have  (See \cite{Landau}, 
Section \S144):
\begin{equation}
 \frac{TA}{2}+ \gamma\Lambda A^{\frac{3}{2}}= -2  {\Phi_0+a(T-T_c)\eta^2+B(P)\eta^4-\eta hA} \label{map1}
\end{equation}
From this, we can now determine the mapping constraint and the dependence of $h$ on $\Lambda$. Using \eqref{eta},  \eqref{map1} 
leads to:
\begin{eqnarray}
 \frac{TA}{2}+ \gamma\Lambda A^{\frac{3}{2}}&=&\left.  -2  [\Phi_0+\kappa a(T-T_c)\frac{A}{2\alpha}+B(P)(\frac{A}{2\alpha})^4 \right. \nonumber\\
 &-&\left.  \frac{1}{\sqrt{\alpha}}\sqrt{\kappa}hA^{\frac{3}{2}}].\right. \label{map2}
\end{eqnarray}
From \eqref{map2}, we can match the coefficients of the different powers of $A$ on both sides. This gives 
\eqref{a} as before and  
\begin{equation}
 h= \sqrt{2\alpha}\frac{1}{\sqrt{\kappa}}\gamma\Lambda. \label{h}
\end{equation}
The mapping constraint remains the same \eqref{2mapc1}. 

In thermodynamic equilibrium, as before, the Thermodynamic Potential should be minimised. That leads to the 
condition,
\begin{equation}
 2a(T-T_c)\eta+4B(P)\eta^3= hA. \label{eqconh}
\end{equation}
Due to the external field $h$, the phase structure system gets richer \cite{Landau}. For $T>T_c$, phase 
transition of the first kind occurs when the system passes through $h=0$, where, phases with $\eta= 
\pm (\frac{-a|T-T_c|}{2B})^{\frac{1}{2}}$, opposite in sign are in equilibrium together. In fact, our analysis
predicts the existence of a tri-critical point \cite{Landau}.  It is interesting to note that in Ref. \cite{Laughlin} 
the black-hole was modeled as a quantum Bose gas near the tri-critical point.

Unlike Schwarzschild black-hole, in this case, the change in the entropy can be performed only in 
two different  --- weak and strong field --- limits. To define the limits, let us denote by
$h_t$, the value of the field at which $\eta_{ind}$ becomes of the same order of magnitude as 
the equilibrium value of the order parameter($\eta_{sp}$) in the absence of any external field($h=0$) 
$\eta_{sp}\sim\frac{(-a|T-T_c|)^{\frac{3}{2}}}{AB^{\frac{1}{2}}}$. [The external field $h$ induces 
change in the order parameter $\eta$. The change in $\eta$ is given by  $\eta_{ind}\sim\chi h$.]
Fields $h\ll h_t$ are ``weak'' as their effect on the thermodynamic variables of the system is small. 
$h\gg h_t$ are strong fields for which the thermodynamic variables have values determined by the field $h$ in 
the first approximation. Thus at $T=T_c$, any field could be treated as strong. 



Let us begin with the case when, $h\ll h_t$ and $h\ll1$. The order parameter $\eta$ could be expressed in 
this case as,
\begin{equation}
 \eta= \eta_0+\eta_1, \label{Weta}
\end{equation}
where, $\eta_0$ is the equilibrium value of the order parameter in the absence of any external field and 
$\eta_1$ is the change in it cause by the presence of an external field. For small values of $h$, the change 
in $\eta$ would be linear in $h$. Hence, using the formulae for susceptibility described earlier, we have 
\begin{equation*}
\eta_1 =
\begin{cases}
 \frac{hA}{2a(T-T_c)} & \text{if } T  <  T_c,\\
\frac{hA}{-4a(T-T_c)} & \text{if } T > T_c .
\end{cases}
\end{equation*}

Using the relation $S= - \partial\Phi/\partial T= S_0-a\eta^2$ and using \eqref{a} and \eqref{eta}, we get, 
$\Delta S = A/4$, same as in the earlier case. Arguing in the same way as in the case of the Schwarzschild 
black hole, we now conclude that the entropy of the black hole system is given by $\frac{A}{4}$ in this case as well. 

AdS-Schwarzschild black hole-fluid system is one for which one can define specific heat. Here $P$ is a 
function of $T$ and 
$\Lambda$. Thus $T$ can be varied keeping $P$ fixed. However, in the weak field limit, where 
$\Lambda\ll1$, $\frac{\Delta Q}{\Delta T}$ is negative as before and the system cannot be in stable 
equilibrium with heat bath.

In the strong field limit, the order parameter in the equilibrium state is given by 
\begin{equation}
 \eta_s= (\frac{hA}{4B})^{\frac{1}{3}}. \label{etas}
\end{equation}
It can be shown in the same way as before that the entropy of the system would be $\frac{A}{4}$. In the 
$h\gg h_t$ limit, the system is a two parameter system. One can define specific heat for this system and it turns out to be \cite{Landau} 
\begin{equation}
 C_P= -\frac{\partial\Phi_0}{\partial T}+ \frac{a^2T_c}{2B}= C_{P0}+ \frac{\alpha^2T_c}{8\kappa^2B}. \label{CPAdS}
\end{equation}
If $C_{P0}$ is positive, then so is $C_P$. Unlike weak field limit, in this case, the system can be in equilibrium with heat bath.

\section{Understanding the order parameter($\eta$)}

In the semi-classical black-hole limit, it is reasonable to assume that 
the fluctuations of $N$ around its equilibrium value goes as, $\Delta N_{r.m.s}\propto\sqrt{N}$, where, 
$\Delta N_{r.m.s}$ is the square root of the mean square fluctuation of $N$.  This implies that,
\begin{equation}
 \Delta N_{r.m.s}\propto|\eta|, \label{etaphys}
\end{equation}
when, $\eta\ne0$. It is important to note that while  $\Delta N_{r.m.s.} \geq 0$, $\eta$ can take 
negative values. Hence, $\eta$ cannot be identified with $\Delta N_{r.m.s.}$.  

A more natural choice would be to identify $\eta$ with the average over the fluctuations in $N$. 
This vanishes when the system is in equilibrium. A non-zero value of this average denotes that the 
system is not in equilibrium. Depending on the sign of the average over the fluctuations, the system is 
driven in a particular direction. Let us now try to quantify this. The constraints \eqref{AE},
\eqref{NA}, 
and the presence of an asymptotic timelike Killing vector ensures that the change in the energy the black 
hole system due to the fluctuations in $N$ must be accounted for. Noting that the Schwarzschild black hole is 
not in equilibrium in the asymmetric phase, we treat the Horizon-Fluid as an open system in contact with an 
external source. There could be an exchange of energy between the two. Due to the constraints, it could also 
be thought of as the exchange of some d.o.f. from one system to the other and vice versa. The fluctuations 
around the mean value of $N$ could then be thought of as having been arisen because of such exchange of 
d.o.f. between the systems. 

Let us assume that the system is at first is in equilibrium with the external source. This is the symmetric 
or the $\eta=0$ phase for a Schwarzschild black hole. The probability of an exchange of a degree of freedom 
from the black hole to the source and its reverse process to take place, when the black hole has $N$ d.o.f.,
is denoted by $\mathcal{P}_{B\rightarrow S}$ and $\mathcal{P}_{S\rightarrow B}$ respectively. At 
equilibrium, $\mathcal{P}_{B\rightarrow S}= \mathcal{P}_{S\rightarrow B}$. This is also the probability that 
the d.o.f. of the black hole increases from the mean value by one due to a fluctuation. Therefore, at 
equilibrium,
\begin{equation}
 \mathcal{P}_{B\rightarrow S}= \mathcal{P}_{S\rightarrow B}\propto\sqrt{N}, \label{}
\end{equation}
where, the average number of d.o.f. of the black hole is given by $N$. 

When not in equilibrium, there would be a net macroscopic exchange of a small number of d.o.f. between the 
black hole and the external source. Let us denote that by $\delta N$. According to our convention, 
$\delta N>0$ ($\delta N<0$), when the black hole gains (loses) in the exchange between the source and the black 
hole. For small exchanges, it is natural to assume the probability that a system loses some d.o.f. would 
depend on the mean number of the d.o.f. of the system only. Let us now consider what happens after the system 
has exchanged $\delta N$ number of d.o.f. between them. We can write,
\begin{equation}
 \mathcal{P}_{B\rightarrow S}\propto\sqrt{N+\delta N}; \mathcal{P}_{S\rightarrow B}\propto\sqrt{N-\delta N}. \label{exprob}
\end{equation}
The average change in the number of black hole d.o.f., $\Delta N_{av}$ would then be given by,
\begin{equation}
 \Delta N_{av}= C \frac{\delta N}{\sqrt{N}}=C\frac{\delta N}{\Delta N_{r.m.s.}}, \label{flucav}
\end{equation}
where, $C$ is some positive constant. So $\Delta N_{av}$ could be positive or negative depending on the sign 
of $\delta N$. 

It is natural then to identify $\eta$ with $\frac{\delta N}{N}$, which would also automatically ensure 
that $\eta\ll1$. Thus we have, 
\begin{equation}
 \eta= \frac{\delta N}{\sqrt{N}}= \frac{\delta N}{\Delta N_{r.m.s.}}=\frac{\delta A}{\Delta A_{r.m.s.}}. \label{etaPhys}
\end{equation}
From \eqref{eta}, then it follows that,
\begin{equation}
 \sqrt{k}= \frac{\delta N}{N}= \frac{\delta A}{A}. \label{k}
\end{equation}

The order parameter $\eta$ thus provides an estimate of how far the black hole is from equilibrium. 
The source mentioned here could be physically realised by some matter-energy that falls into the 
Schwarzschild black hole. 
Typically the energy of such infalling matter would be much smaller than that of the black hole and 
$\frac{\delta N}{\Delta N_{r.m.s}}$, would typically scale by some power of the Planck length, also a tiny 
number. For an observer far from the black hole horizon, any such infalling matter reaches the horizon 
asymptotically. Such an observer sees the area of the black hole horizon increasing quasi-statically due to 
the matter influx. The phase characterized by $\eta<0$ could be realised for Hawking radiation. The external 
source in this case is the thermal radiation. The black hole emits more radiation and its area decreases.

\section{Discussion}

Our analysis allows us to draw an important conclusion.  {\it The Schwarzschild black hole has 
an entropy given by the Bekenstein-Hawking formula only when it is not in equilibrium with its surroundings.}
We expect this result to hold for Kerr and Reisner-Nandstorm black holes, away from the extremal limit. 
Due to spontaneous symmetry breaking, the black hole could go into any one of the phases
($\eta>0$ and $\eta<0$) with equal likelihood. The formalism developed here is also not suitable to conclude 
whether, in the non-equilibrium phases, the black hole would exhibit runaway behaviour or not. Considerations about the 
change in entropy is useful here. $\frac{dS}{d\eta}>0$, only if $\eta>0$. Hence, 
entropy-wise, it is favourable for a black hole in the $\eta>0$ phase to go on increasing its area by 
absorbing more matter. This tendency is the cause of the out of equilibrium and negative specific heat of the 
black hole \cite{HP}. In $\eta<0$ phase, the horizon area would continue to decrease if the decrease 
in the area and the entropy of the black hole is compensated by an increase in the entropy of the radiation 
outside the black hole, so that the total entropy actually increases \cite{HP}. This happens in the emission 
of Hawking radiation\cite{Zurek} and in this case also, the black hole shows runaway behaviour.  

The negative cosmological constant can also be thought of as an external field which induces 
order in the horizon fluid. Recalling \eqref{h}, we see that this, however, is only possible if, 
$\delta N\neq0$. For ADS-Swarzschild black holes, the phases with $\eta>0$ and $\eta<0$ 
could coexist. According to the considerations discussed above, these are the phases with classical and 
quantum instabilities respectively. But the amount of matter-energy exchanged between the black hole and its 
surroundings would be opposite in sign. So, if we consider, the system as a whole, the net exchange between 
the black hole  and its surroundings could be zero. Then the black hole would be in thermal equilibrium with 
its surroundings. This is seen from the specific heat becoming  positive from a negative value, for large values of $\Lambda$. 
However, it is possible to have a stable horizon fluid system only for a negative cosmological constant. Our 
analysis suggests, that a black hole in an asymptotically de-Sitter space might be 
unstable\footnote{This is because, in an asymptotically de-Sitter space, a cosmological horizon with a 
temperature different
than the black hole horizon would also be present.}. For the AdS-Schwarzschild black hole, the system has a 
phase diagram with a tri-critical point. This 
relates to some other attempts made in the literature to describe the black hole physics by the physics near 
a critical point\cite{Laughlin}. What is remarkable here is that, it is possible to give an explanation for
these well-known features of Black Hole Thermodynamics from the horizon fluid perspective. 

As is well known, Mean Field Theory is valid only when the fluctuations of the order parameter are 
irrelevant. However, in this work, we have shown that, such a crude approximation, in particular in 
2D\cite{Goldenfeld}, could lead to Bekenstein-Hawking entropy. Our work strongly suggests 
that any approach that predicts Bekenstein-Hawking entropy should be treated at the same level as a Mean 
Field model in Condensed matter systems that implicitly ignore fluctuations. It would be interesting to see 
whether mean field theory can predict the change in the entropy for higher-derivative gravity theories.  Our 
work also suggests that going beyond Mean Field Theory will lead to fundamental understanding of black hole 
entropy. Finally, for smaller black holes having a larger temperature, the low energy theory having $Z_2$ 
symmetry may no longer be valid. Then, there is a possibility, that something like the Kosterlits-Thouless 
transition takes place \cite{Domany2} as the temperature of the black hole increases, i.e. as it gets 
smaller.

\section*{Acknowledgments}

The work is supported by Max Planck-India Partner Group on Gravity and
Cosmology. SS is partially supported by Ramanujan Fellowship of DST,
India.

\begin{center}
\section*{Appendix}
\end{center}

\appendix

\section{Symmetry Breaking: $U(1)$ to $Z(2)$ via RG Flow}\label{RG}

We start with the general form of the Landau-Ginzberg Theory for a system describing a BEC. The order 
parameter is denoted by $\psi$ in this case, which is could take complex values. The energy of the system 
is given by, 
\begin{equation}
 E= \int (|\nabla\psi|^2-\mu|\psi|^2+g|\psi|^4) dV. \label{LGInh}
\end{equation}
The partition function for this Theory is given by,
\begin{equation}
 Z= \int D\psi\exp[-\beta( \int [|\nabla\psi|^2-\mu|\psi|^2+g|\psi|^4] dV)]. \label{partpsi}
\end{equation}

Let us now assume that the phase part of the order parameter comes from the high energy d.o.f. of the 
system. Then the fluctuations in the phase part of the order parameter are characterized by a much smaller 
length scale compared to the scale characterizing the fluctuations in the real part of $\psi$. Hence, one may 
integrate out the phase d.o.f. if one is interested only in a low energy description of the system. To do 
this, we split $\psi$ into two parts, via 
\begin{equation}
 \psi= \eta+\tilde{\psi}, \label{FS}
\end{equation}
where, $\eta$ is real valued and $\tilde{\psi}$ is complex. According to the assumption made by us, 
$\tilde{\psi}$ varies much faster than $\eta$. Therefore, we may write, 
\begin{equation}
 |\nabla \tilde{\psi}|\gg1, \label{fastvar}
\end{equation}
where, we have normalised, so that, $\nabla\eta \sim1$. So we can write the partition function as 
$Z(\psi)= Z(\eta, \tilde{\psi})$.  It can be expressed as a path integral given by 
\begin{equation}
 Z(\psi)= \int D\eta D\tilde{\psi} e^{-\beta E(\eta, \tilde{\psi})}. \label{partition1}
\end{equation}
Now we expand $E(\eta+ \tilde{\psi})$ around $\eta$ with respect to $\tilde{\psi}$. This gives,
\begin{equation}
 Z(\psi)= \int D\eta e^{-\beta[\int[(\nabla\eta)^2-\mu\eta^2+g\eta^4]dV]}D\tilde{\psi} e^{-\beta E_{rest}(\eta, \tilde{\psi})}. \label{partition2}
\end{equation}
At this point, we could gain more insight into this path integral by noting the following facts. Firstly, 
$|\psi|\ll1$ . The system is in thermodynamic equilibrium, when $\psi$ satisfies the equation, 
$\frac{\delta E}{\delta\psi}=0$, which could be referred to as the stationary condition. Now, though in 
general, $\psi$ takes complex values, it would also take real values. So the space of all possible values of 
$\psi$ that satisfy the stationary condition would also include real numbers. The path integral in 
\eqref{partition2} would include real values of $\psi$ that satisfy the stationary condition. These would 
also be some of the possible values that $\eta$ could take and and we are expanding $E$ around these values 
with respect to $\tilde{\psi}$. Let us denote these stationary values of $\eta$ by $\eta_s$. Now let us 
expand $Z$ around $\eta_s$ and we define a new variable, $\delta\eta$ by the relation, 
$\delta\eta= \eta-\eta_s$. Also, $\eta_s=\eta_s(\mathbf{x})$. Then one can rewrite \eqref{partition2} as,
\begin{widetext}
\begin{equation}
 Z(\psi)= \sum\limits_{\mathbf{\eta_s(x)}}\int D\delta\eta e^{-\beta[\int[(\nabla\delta\eta)^2-\mu\delta\eta^2+g\delta\eta^4]dV]}D\tilde{\psi} e^{-\beta E_{rest}(\delta\eta, \tilde{\psi})}. \label{partition3}
\end{equation}
\end{widetext}
 Because of the stationary condition, the term linear or anti linear in $\psi$ would be zero. The leading 
 order potential term in the expansion could then either be 
proportional to $\delta\eta^2$ or $|\tilde{\psi}|^2$. The $\delta\eta^2$ term has already been taken care of 
in the part of the path integral over $\delta\eta$ configurations. Then the leading order potential term in 
the expansion of $E$ around $\eta_s$ with respect to $\tilde{\psi}$ is the term proportional to 
$|\tilde{\psi}|^2$. In a similar way, it can be argued that the leading order kinetic term in the expansion 
could only be proportional to $|\nabla\tilde{\psi}|^2$. Then we can write the partition function as,
\begin{widetext}
\begin{equation}
 Z(\psi)=\sum\limits_{\mathbf{\eta_s(x)}}\int_{\delta\eta_{Configs}} [D\delta\eta e^{-\beta[\int [(\nabla\delta\eta)^2-\mu(\delta\eta)^2+g(\delta\eta)^4]dV]}\int_{\tilde{\psi}_{Configs}}D\tilde{\psi} e^{-\beta[\int (C|\nabla\tilde{\psi}|^2+ D|\tilde{\psi}|^2+higher\ order\ terms\ in |\tilde{\psi}|)dV]}] \label{partition4} 
\end{equation}
\end{widetext}
\footnote{$\eta_s$ satisfies the condition of being a minima of the potential as well, 
except for the point $\eta_s=0$, where, $\frac{\delta^{(n)} E}{\delta \psi^{(n)}}=0$ for $\forall n$. But the 
the measure of the path integral in terms of the variable expanded around $\eta_s=0$ is zero. Hence, this 
point can be included in the expression for $Z$, as it has no contribution.}, where, we have 
used the $U(1)$ symmetry of the Energy function. 

To use the condition \eqref{fastvar}, one has to switch over to a momentum space representation of 
$\tilde{\psi}$, given by,
\begin{equation}
  \tilde{\psi}= \frac{1}{(2\pi)^\frac{n}{2}}\int_{-\infty}^{+\infty}\tilde{\Psi}_{\mathbf{k}} e^{i\mathbf{k}.\mathbf{x}}dV. \label{FT}
\end{equation}

\begin{widetext}
 \begin{equation}
   Z(\psi)=\int_{\delta\eta \space Configs} [D\eta e^{-\beta[\int [(\nabla\delta\eta)^2-\mu(\delta\eta)^2+g(\delta\eta)^4]dV]}\int_{\tilde{\Psi}_{\mathbf{k}}\space Configs}\prod\limits_{\mathbf{k}} D\tilde{\Psi}_{\mathbf{k}} e^{-\beta(\int (k^2 C_{\mathbf{k}}+D_{\mathbf{k}})|\tilde{\Psi}_{\mathbf{k}}|^2+ higher\ order\ terms)d^nk)}]. \label{partition5}
 \end{equation}
\end{widetext}
\eqref{fastvar} implies that $k^2\gg1$. This ensures that we could replace the entire integrand by the 
$|\tilde{\Psi}_{\mathbf{k}}|^2$ term in the path integrals of the form 
$\int D\tilde{\Psi}_{\mathbf{k}}f(\tilde{\Psi}_{\mathbf{k}})$  in the r.h.s. of \eqref{partition5} as a good 
approximation. Then the 
partition function could be written as 
\begin{widetext}
\begin{equation}
 Z(\psi)\approx \int_{\delta\eta \space Configs} [D\delta\eta e^{-\beta[\int [(\nabla\delta\eta)^2-\mu(\delta\eta)^2+g(\delta\eta)^4]dV]}\int_{\tilde{\Psi}_{\mathbf{k}}\space Configs}\prod\limits_{\mathbf{k}} D\tilde{\Psi}_{\mathbf{k}} e^{-\beta(\int (k^2 C_{\mathbf{k}}+D_{\mathbf{k}})|\tilde{\Psi}_{\mathbf{k}}|^2d^nk)}]. \label{partition6}
\end{equation}
\end{widetext}
The phase d.o.f. or $\tilde{\psi}$ could now be integrated out of $Z$ by performing Gaussian path integrals 
in $\tilde{\Psi}_{\mathbf{k}}$. This would result in a constant factor, that would be denoted 
here by $\frac{1}{\Delta}$. It could be taken out of the path integral. Hence, one gets 
\begin{equation}
 Z(\psi)= \frac{1}{\Delta} \int D\delta\eta e^{-\beta[\int [(\nabla\delta\eta)^2-\mu(\delta\eta)^2+g(\delta\eta)^4] dV]}. \label{partition7}
\end{equation}
Making a change of variable from $\delta\eta$ to $\eta$, one gets the partition function to be of the same 
form as given by \eqref{partition7}, modulo some constant factor. Ignoring the constant factors, the 
partition function can be written as 
\begin{equation}
 Z(\eta) =  \int D\eta e^{-\beta\int [(\nabla\eta)^2-\mu\eta^2+g\eta^4]dV}, \label{partitioneta}
\end{equation}
the partition function describing the low energy physics of the same system. This low energy theory is 
governed by the form of energy expressed only in terms of the low energy d.o.f. and is given by,
\begin{equation}
 E_{\eta}= \int [(\nabla\eta)^2-\mu\eta^2+g\eta^4]dV. \label{Elow}
\end{equation}

This can now be treated as a theory which has a $Z2$ symmetry, a discrete symmetry. The spontaneous symmetry 
breaking that would occur when the energy gets minimised would be a $Z2$ symmetry breaking, As an aside, we 
note here that some minor modifications of this technique would give us the same result in Field Theory.

\section{2-D Ideal massless Gas}
The Einstein equations projected on the event horizon of a Schwarschild black hole could be described by a 
$2+1$ dimensional relativistic fluid that resides on the event horizon of the black hole \cite{Damour}. The 
volume of the fluid is the area of the horizon, denoted by $A$. The temperature of the horizon, denoted henceforth by $T$, 
is the temperature of the fluid and the total energy of the fluid is given by the Komar mass of the black 
hole \cite{Skakala}. The equation characterizing the fluid is then given by \cite{Skakala},
\begin{equation}
 P= \frac{T}{4}=\frac{E}{2A}. \label{eqs}
\end{equation}
This is the equation of state of a $2D$ ideal massless relativistic gas \cite{Huang}.

If the number of degree of freedom in the horizon-fluid system is given by $N$, then the entropy is typically 
a function of three parameters, $S(E,N,A)$. However, the parameter space of the Schwarzschild black hole is 
one dimensional. So $E$, $N$, $A$ are not independent, but must obey the two constraints: $E= E(A)$ and 
$N=N(A)$. The constraint equation between $E$ and $A$ is given by\cite{Skakala}, 
\begin{equation}
 A = 16\pi E^2. \label{AE}
\end{equation}
 The constraint relating $N$ and $A$ could be derived from  the equation relating the black hole mass and the
 Hawking temperature, 
\begin{equation}
 E= \frac{1}{8\pi T}. \label{ET}
\end{equation}
Deriving $N(A)$ from \eqref{ET}, however, requires information about the statistical model and would be done 
at a later stage.

\subsection{The Statistical Physics Viewpoint} 
From the preceding discussion, it is clear that the fluid in the fluid-horizon correspondence in Damour's 
work \cite{Damour}, could be viewed as a collection of microscopic degrees of freedom, which obey Bose 
statistics. In \cite{Skakala}, the authors take them to be a collection of $N$ particles.  As described in 
\cite{Skakala}, the energy levels of a free relativistic  
particle  could be calculated. They also give the spectrum of a massless relativistic scalar particle living 
on the surface of a sphere of area $A$ \cite{Kleinert}, 
\cite{Skakala},
\begin{equation}
\epsilon^r_l= \sqrt{\frac{4\pi{l(l+1)+\alpha^2}}{A}}= \tilde{\epsilon}_lT, \label{spectrum}
\end{equation}
where, $\tilde{\epsilon}_l= \sqrt{l(l+1)+\alpha^2}$. $\tilde{\epsilon}_l$ is defined in such a way that it is 
independent of the black hole parameters. 

Here we argue 
on general grounds that for a microscopic degree of freedom residing on a spherical surface of area $A$, 
the spectrum would be given by a form similar to that in \eqref{spectrum}. If we assume that the microscopic 
degrees of freedom on the horizon are independent of the physics in the bulk, then the only length scale for 
them is the one set by the area of the horizon. On dimensional grounds then, one can write, the energy 
levels of such a microscopic degree of freedom in the following way, $\epsilon \propto \frac{1}{\sqrt{A}}$.
Because of the black hole constraints, this could be expressed as $\epsilon \propto T$. This would give an 
expression for the energy levels similar in form to the expression for the energy levels of a particle given 
by \eqref{spectrum}. Then one can express the energy of the ground state of such a microscopic degree of 
freedom as, $\epsilon_0 = \alpha T$, where, $\alpha$ is a constant. It is to be noted that, very little input 
about the microscopic degrees of freedom has gone into our analysis so far. All that has been assumed is the 
existence of $N$ such degrees of freedom and their Bosonic nature.

 \section{Schwarzschild AdS: Black hole-fluid correspondence}
 Let us denote the horizon radius of the black hole by $r_h$. Then the area $A$ is given by 
 $A= 4\pi r_h^2$. 
 Let us denote the cosmological constant by $\bar{\Lambda}$. Then, we define $\Lambda=-\bar{\Lambda}$.
 Using the expression for the Komar mass for an AdS- Schwarzschild black hole \cite{ADSBH}, the energy of the 
 horizon-fluid system, 
 \begin{equation}
  E= \frac{TA}{2}+ \frac{1}{48G(\pi)^{\frac{3}{2}}}\Lambda A^{\frac{3}{2}}. \label{AdSE}
 \end{equation} 
 The pressure of the black hole horizon-fluid is given by\cite{ADSBH},
 \begin{equation}
  P=\frac{E}{2A}= \frac{T}{4}+\gamma\Lambda\sqrt{A}, \label{AdSP}
 \end{equation}
 where, 
  \begin{equation}
  \gamma= \frac{1}{48G(\pi)^{\frac{3}{2}}}. \label{gamma}
 \end{equation}
 \eqref{AdSP} is the equation of state of the AdS-Schwarzschild black hole-fluid.  
 \eqref{AdSE} and \eqref{AdSP} could then be written as 
 \begin{equation}
  E=A(\frac{T}{2}+\gamma \Lambda\sqrt{A}) \label{AdSE1}
 \end{equation}
 and 
 \begin{equation}
  P= \frac{T}{4}+ \gamma\Lambda\sqrt{A}. \label{AdSP1}
 \end{equation}

 The constraint equations in this case are different from that in the case of a Schwarzschild black hole 
 because of the presence of $\Lambda$. The relation between the area and the mass of the black hole can be 
 found from the relation
 \begin{equation}
  \frac{1}{3}\Lambda r_h^3- r_h+2M=0 \label{AdSMA}
 \end{equation}


\begin{thebibliography}{100}
\bibitem{BH-Thermo} 
J.D. Bekenstein, Phys. Rev. {\bf D7}, 2333 (1973); 
%
J.M. Bardeen, B. Carter and S.W. Hawking, Comm. Math. Phys. {\bf 31}, 161 (1973);
%
Carter, B., in {\sl General Relativity: An Einstein Centenary Survey}, editors
S.W. Hawking and W. Israel, Cambridge U. Press, (1979).

\bibitem{Hawking-1975}
S.W. Hawking, Commun. Math. Phys. {\bf 43}, 199 (1975).

\bibitem{Shiraj} 
S.~Bhattacharyya, S.~ Minwalla, V.~E.~Hubeny, M.~Rangamani, JHEP {\bf 045}, 0802 (2008)

\bibitem{Paddy} 
T.~Padmanabhan, Phys. Rev. D, {\bf 83}, 044048 (2011);
S.~Kolekar, T.~Padmanabhan, Phys. Rev. D, {\bf 85}, 024004 (2011).
 
\bibitem{Strominger} 
I.~Breedberg, C.~Keeler, V.~Lysov, A.~Strominger, JHEP {\bf 2017} 146, (2012);
I.~Breedberg, A.~Strominger, JHEP {\bf 1205}, 043, (2012).
  
\bibitem{Sakharov} A.~D.~Sakharov, Sov. Phys. Dokl. {\bf 12} 1040-1041 (1968) \\
 (Translated) GRG, {\bf 32} 365-367 (2000).
 
 \bibitem{Damour} T.~Damour, {\it Surface effects of black hole physics}, Proc. of M. Grossman Meeting 
 (1982), North Holland, p.587.
 
 \bibitem{Membrane} R.~H.~Price, K.~S.~Thorne, Phys. Rev. D, {\bf 33}, 915 (1986);
 K.~S.~Thorne, D.~A.~ MacDonald, R.~H.~Price, {\it Black Holes: The Membrane Paradigm (The 
 Silliman Memorial Lectures Series)}; Yale University Press (1986).
 
\bibitem{Laughlin} G.~Chapline, E.~Hohlfeld, R.~B.~Laughlin, D.~I,~Santiago, Int.J.Mod.Phys. A, {\bf 18}, 
 3587-3590 (2003). 
 
 \bibitem{Dvali} 
  G.~Dvali, C.~Gomez, Phys. Lett. B, {\bf 716}, 240 (2012);  Fortcshr. Phys. {\bf 63}, No 7-8, 742-767 (2013); 
 Eur. Phys. J. C, {\bf 74}, 2752 (2014).
 
  \bibitem{stretchrzn} S.~Zare, Z.~Rasssi, H.~Mohammadzadeh, B.~Mirza, Eur. Phys. J. C,, {\bf 72}, 
  2152 (2012). 
 
 \bibitem{Skakala} J.~Skakala, S.~Shankarnarayanan, arxiv: 1406.2477.
 
 \bibitem{Mermin} N.~D.~Mermin, H.~Wagner, Phys. Rev. Lett. {\bf 22}, 17,  1133 (1966).
 
 \bibitem{Hohenberg} P.~C.~Hohenberg, Phys. Rev. {\bf 2}, 158, 383 (1967).
 
 \bibitem{KadanoffKane} J.~W.~Kane, L.~P.~Kadanoff, Phys. Rev. {\bf 1}, 155, 80 (1967) 
 
 \bibitem{Nature} Z.~Hadzibabic, P.~Krug\"{e}r, M.~Cheneau, B.~Battelier, J.~Dalibard, Nature Lett., 44, 
 1118 (2006) 

 \bibitem{Cho} W.~Cho, S.~Kim, J.~Park, arxiv: 1409.4277[cond-mat.quanta-gas](2014).
 
 \bibitem{Huang}K. Huang, {\it Statistical Mechanics}, John Wiley \& Sons, 1987.
 
 \bibitem{Kleinert}H.~Kleinert, S.~Shabanov, Phys. Lett. A, {\bf 232}, 327-332 (1997). 
 
 \bibitem{Landau} L.~D.~Landau \& E.~M.~Lifshitz, {\it Statistical Physics, Part 1}; Butterworth-Heinemann 
 (1980). 
 
 \bibitem{Stanley} H.~E.~Stanley, {\it Introduction to phase transitions and critical phenomena}; Oxford 
 University Press (1971).
 
 \bibitem{Kadanoff} L.~P.~Kadanoff, G.~Baym, {\it Quantum Statistical Mechanics};
 Westview Press (1994); L.~P.~Kadanoff, {\it Statistical Physics; Statics, Dynamics and Renormalization}; 
 World Scientific Publishing (2010)
 
 \bibitem{Domany1} E.~Domany, E.~K.~Riedel, Journal of App. Phys., {\bf 49}, 1315, (1978).
 
 \bibitem{Rottman} C.~Rottman, Phys. Rev. B, {\bf 3}, 24, (1981).
 
 \bibitem{Domany2} E.~Domany, B. Mutaftschiev(ed.), {\it Interfacial Aspects of Phase Transformation}, 
 119-141, D. Reidel Publishing Company, (1982). 
 
 \bibitem{ADSBH} M.~Cveti\u{c}, G.~W.~Gibbons, D.~Kubiz\v{n}\'{a}k, C. ~N.~Pope, Phys. Rev. D, {\bf 84}, 
 024037 (2011).  
 
 

 \bibitem{HP} S.~W.~Hawking, D.~N.~Page, Commun. Math. Phys., {\bf 87}, (1983), 577-588.
 
 \bibitem{Zurek} W.~H.~Zurek, Phys. Rev. Lett,, {\bf 23}, 49, (1982), 1683.
 
 \bibitem{Goldenfeld} N.~Goldenfeld, {\it Lectures On Phase Transitions and The Renormalization Group}, 
 (2005) Levant Books, Kolkata.
\end{thebibliography}
\end{document}